\documentclass{stavanger}

\usepackage[utf8]{inputenc}
\usepackage{amssymb,amsmath,amsfonts,amsthm}
\usepackage{hyperref}
\usepackage{natbib}
\usepackage{graphicx}
\usepackage{tikz}
\usepackage{cleveref}
\usepackage{ulem} 
\usepackage{placeins}

\startlocaldefs

\endlocaldefs

\begin{document}


\begin{frontmatter}

\begin{fmbox}


\dochead{Research Article - Preprint}{FP}


\title{Conserving Lattice Gauge Theory for Finite Systems}


\author[
   addressref={aff1},                   	  
   corref={aff1},                     		  
   email={alexander.rothkopf@uis.no}   		  
]{\inits{AR}\fnm{Alexander} \snm{Rothkopf}}


\address[id=aff1]{
  \orgname{Faculty of Science and Technology}, 	 
  \street{University of Stavanger},                     		 
  \postcode{4021}                               			 
  \city{Stavanger},                              				 
  \cny{Norway}                                   				 
}



\end{fmbox}


\begin{abstractbox}

\begin{abstract} 
In this study I develop a novel action for lattice gauge theory for finite systems, which accommodates non-periodic boundary conditions, implements the proper integral form of Gauss' law and exhibits an inherently symmetric energy momentum tensor, all while realizing automatic ${\cal O}(a)$ improvement. Taking the modern summation-by-parts formulation for finite differences as starting point and combining it with insight from the finite volume strategies of computational electrodynamics I show how the concept of a conserving discretization can be realized for non-Abelian lattice gauge theory. Major steps in the derivation are illustrated using Abelian gauge theory as example.
\end{abstract}



\end{abstractbox}


\end{frontmatter}


\section{Introduction}

Revealing the dynamical properties of matter under extreme conditions, be it in the context of ultra-high temperature and density, or ultra-strong interparticle coupling, lies at the heart of modern theoretical physics. Non-perturbative access to the quantum many-body physics of such systems is called for and lattice gauge theory has established itself as a vital tool \cite{detmold_topical_2019}. 
It provides the conceptual and computational basis to predict phenomenologically relevant observables over a vast range of energies, one pertinent example being lattice Quantum-Chromodynamics (QCD). Lattice QCD predicts from first principles not only the properties of stable hadrons (see e.g.~\cite{gpd_lat_2020,borsanyi_ab_2015}) and resonances \cite{briceno_scattering_2018} but captures the thermodynamic properties of a strongly interacting many-body system, such as the quark-gluon plasma (QGP) \cite{bazavov_hot-dense_2019,axion_lat_2016}. On the other hand at lower energies, recent experimental advances in the manipulation of confined light fields have made accessible a regime of non-perturbatively large couplings in cavity quantum electrodynamics (QED) (for recent reviews see \cite{cQED1_2019,cQED2_2019}), which defies the standard methods of perturbation theory. A lattice field theoretical approach to this deep-strong coupling (DSC) regime, in particular, in the presence of large number of interacting dipoles promises urgently needed insight (see also \cite{jaako_quantum_2020}) into an as of yet unexplored realm.  
 
In each case, the study of finite quantum systems and the physical consequences of non-trivial boundary conditions play a central role. At high temperatures, the unexplained signatures of collectivity in small collision systems (see \cite{nagle_small_2018} for a review) urge theory to provide non-perturbative insight into strongly-interacting many-body systems of finite extent (see e.g. \cite{mogliacci_geometrically_2020,Panero:2008mg}). At low energies the confines of the reflective cavities used to manipulate the photon-atom interactions leaves an essential imprint on spectra both on the single particle level \cite{casanova_deep_2010} as well as on collective excitations. In order to make decisive progress in these fields, non-perturbative theory support is essential, which however can only be provided by a proper lattice gauge theory for finite systems.

According to Wilsons seminal work \cite{wilson_confinement_1974}, the simplest gauge invariant action for non-Abelian fields on a lattice with periodic boundary conditions (PBC), spatial and temporal lattice spacings $a_s$, $a_t$ and grid points $N_s$, $N_t$ can be expressed \cite{klassen_anisotropic_1998} as $S=\frac{1}{g^2}\sum_{x}\big[ \sum_i 2\frac{a_s}{a_t} {\rm ReTr} [P_{0i}-1] -\sum_{i,j}\frac{a_t}{a_s} {\rm ReTr} [P_{ij}-1]\big]$, in terms of link variables $U_{\mu,x}={\rm exp}[i a_\mu A_{\mu,x}]$, combined in elementary plaquettes
\begin{align}
\nonumber P^{1\times1}_{\mu\nu,x}&=U_{\mu,x} U_{\nu,x+a_\mu{\hat\mu}}U^\dagger_{\mu,x+a_\nu{\hat\nu}}U^\dagger_{\nu,x} = e^{i a_\mu a_\nu \tilde F_{\mu\nu,x}}+{\cal O}(a^2),\\
&\tilde F_{\mu\nu}=\Delta^{\rm F}_\mu A_{\nu,x} - \Delta^{\rm F}_\nu A_{\mu,x} + i [ A_{\mu,x},A_{\nu,x}].\label{eq:Waction}
\end{align} The forward finite differences (FD) are defined as $\Delta^{\rm F}_\mu \phi(x)= (\phi(x+a_\mu \hat\mu) - \phi(x) )/a_\mu$ and the generators $T^a$ of the gauge group enter as $A_{\mu,x}=A^a_{\mu,x}T^a$. 

The past decades have seen significant progress in the formulation of lattice gauge theory beyond \cref{eq:Waction}. Symanzik's improvement program \cite{symanzik_continuum_1983} laid out how to systematically reduce discretization errors by adding higher dimensional operators, leading to highly improved actions, such as in \cite{GarciaPerez:1993lic,Beinlich:1995ik,deForcrand:1997esx,Snippe:1997ru,BilsonThompson:2002jk,Langfeld:2004tf}. On the other hand optimized algorithms exploit the structure of such actions to realize efficient Monte-Carlo simulations of the corresponding Euclidean path integral \cite{Jansen:2008vs}.

\section{Challenges in finite systems}

\begin{figure} [h] 
\includegraphics[scale=0.3,clip=true, trim= 0 0 8.28cm 0 ]{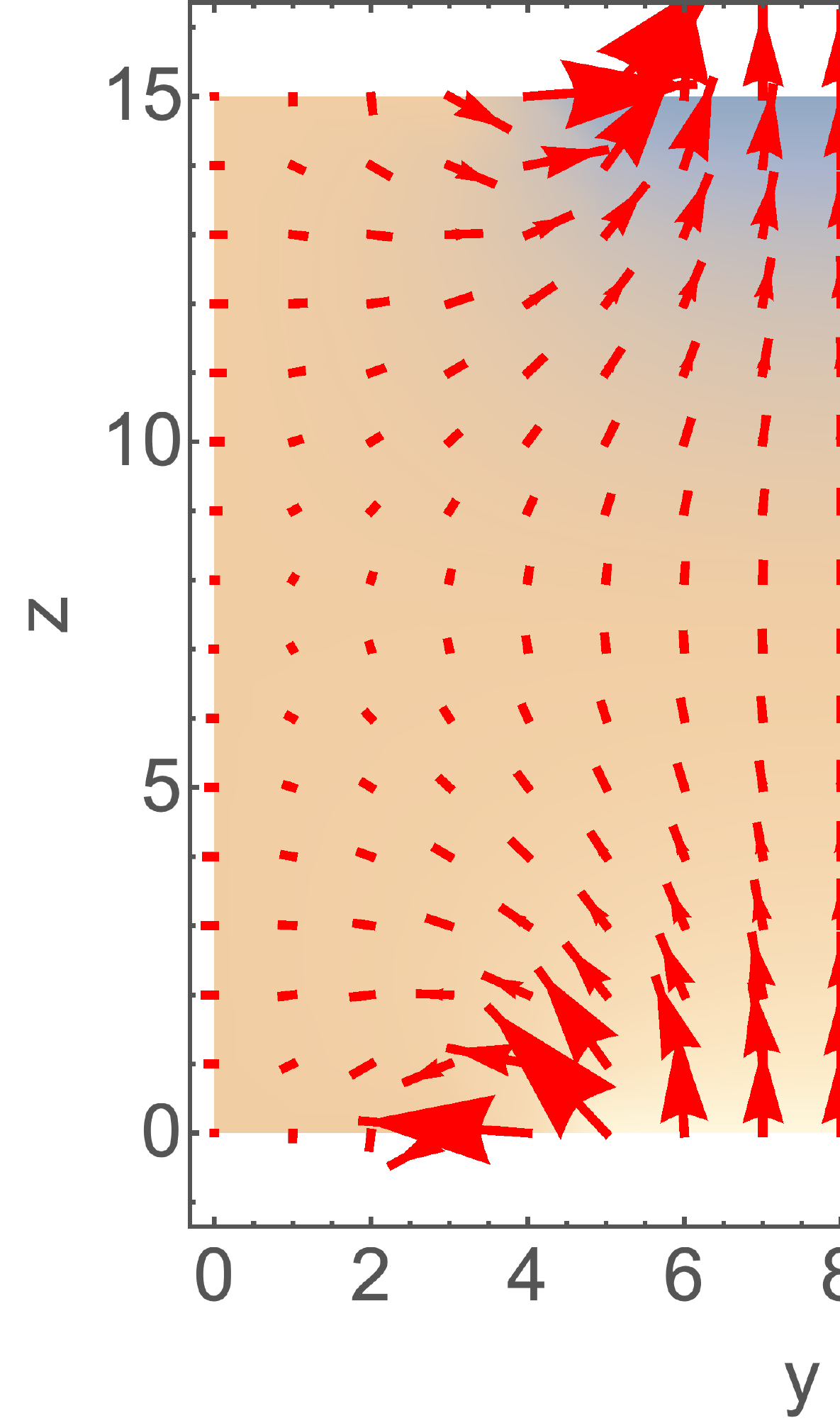}\includegraphics[scale=0.3,clip=true, trim= 12.02cm 0 0 0 ]{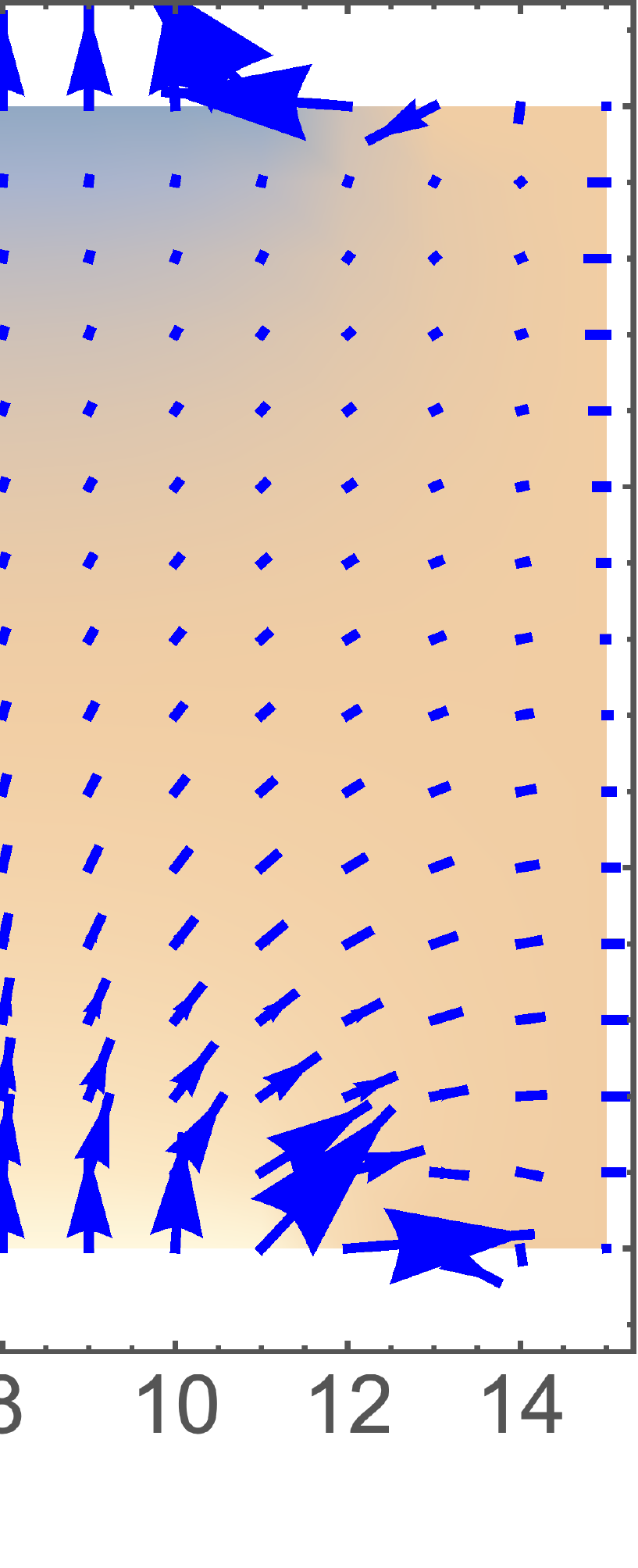}\includegraphics[scale=0.3,clip=true, trim= 2.6cm 0 8.28cm 0 ]{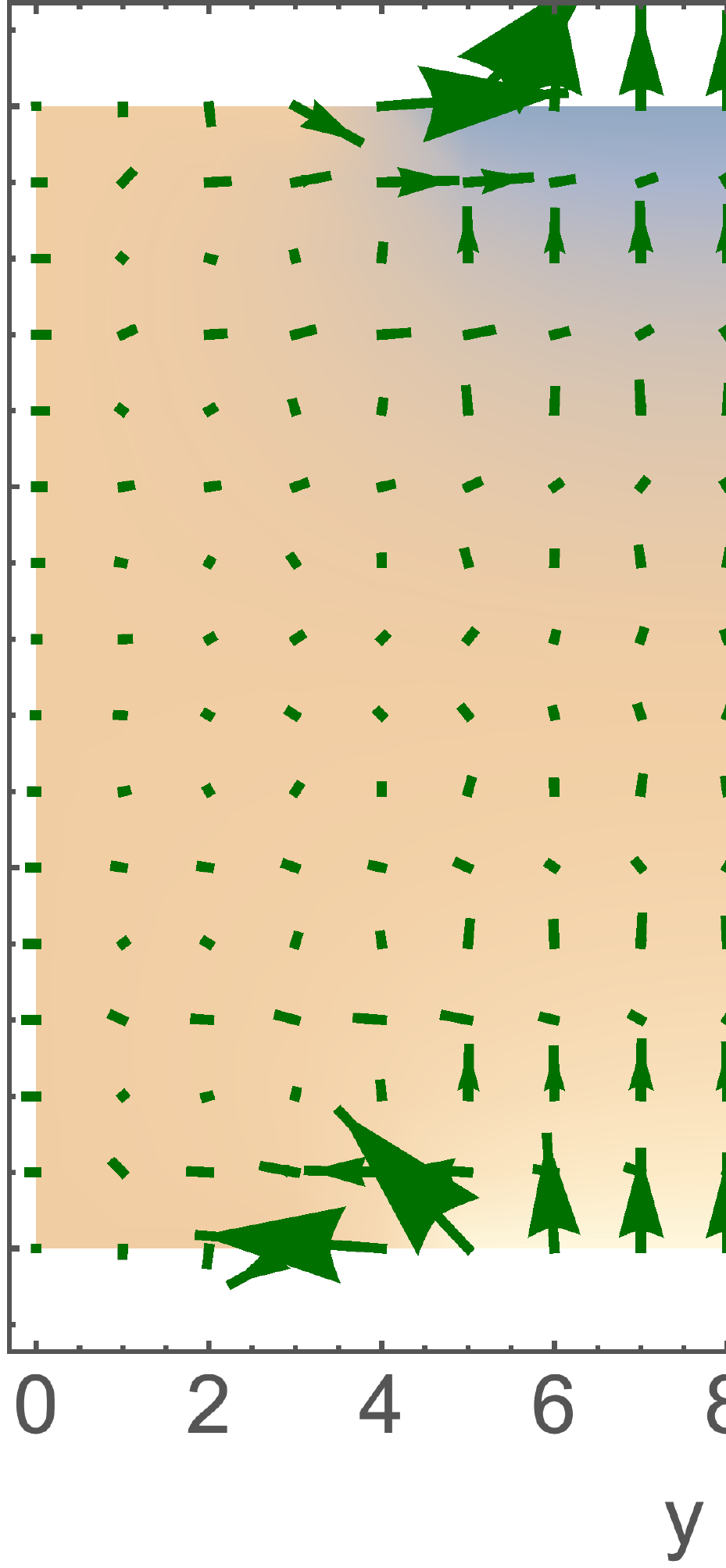}\includegraphics[scale=0.3,clip=true, trim= 12.02cm 0 0 0 ]{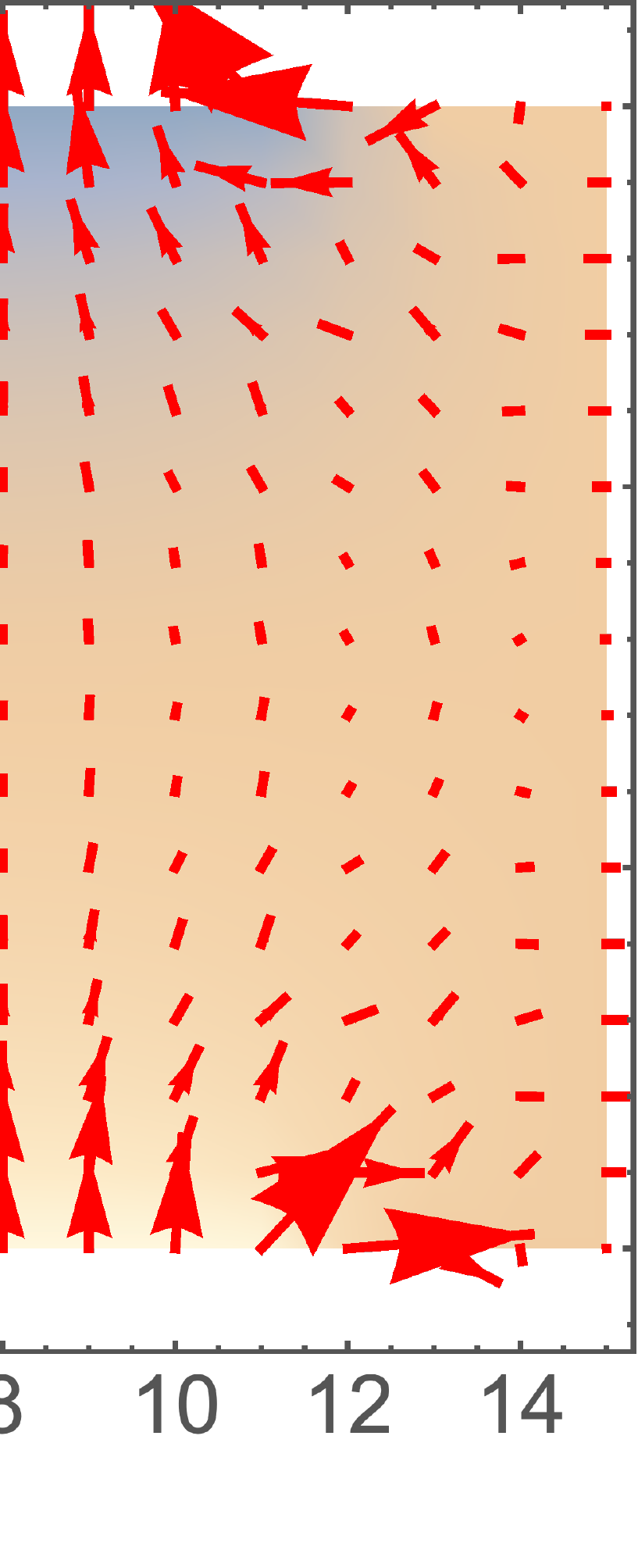}
\caption{(left) Electric field (red arrows) from the capacitor potential (background) vs. backward FD solution (blue arrows). (right) The field strength from a naive central FD (green arrows) and the more accurate solution based on a SBP discretization (red arrows). } \label{fig:CapacitorExample}
\end{figure}

Two challenges exist in deploying the discretization of \cref{eq:Waction} in a finite system, where PBC are inapplicable and translational invariance is broken: (A) a naive first order FD scheme does not accurately reproduce the true solution of a differential equation if it remains restricted to the interior of the solution domain. Let us take a look at the Abelian Gauss law ${\bf \nabla}\cdot {\bf E}=0$ in the interior of a finite circular capacitor, whose end-plates are located on the boundary in the z-direction, held at a finite potential. The accurate solution based on the electric potential in the centered yz plane is shown as red arrows in the left half of the left panel of \cref{fig:CapacitorExample}. Setting this solution as Dirichlet boundary for a naive backward FD approximation ${\bf \Delta}^{\rm B}\cdot {\bf E}=0$ we obtain the blue arrows instead. The approximation is unable to access the forward facing boundary and thus cannot sustain field strength close to it. The result improves when using a naive central FD $\Delta^{\rm C}_\mu \phi(x)= (\phi(x+a_\mu \hat\mu) - \phi(x-a_\mu \hat\mu) )/2a_\mu$ in ${\bf \Delta}^{\rm C}\cdot {\bf E}=0$. As seen from the green arrows in the right panel of \cref{fig:CapacitorExample}, field strength is generated close to the forward boundary but the field lines show an artificial staggered pattern.

The correct treatment of non-trivial boundary conditions in FD approximations is reflected in their implementation of the discrete counterpart to integration by parts, the so called summation by parts (SBP) property (for recent reviews see \cite{svard2014review,fernandez2014review}). Using the trapezoid rule for integration $\int_0^L dx f(x)g(x) \approx {\cal T}_0^N[f_xg_x]$ , one must require the first order FD approximation to fulfill
\begin{align}
  {\cal T}_0^N[ (\Delta^{\rm SBP}f_x)g_x]   \overset{!}{=} -{\cal T}_0^N[ f_x (&\Delta^{\rm SBP}g_x)] + f_{N}g_{N}- f_0g_0.
\end{align}
The lowest order implementation of such an SBP operator is obtained by combining the central FD stencil in the interior with the forward FD and backward FD stencil at the boundaries. The resulting operator is anti-symmetric (corresponding to a hermitean momentum operator on the whole domain) and when used in ${\bf \Delta}^{\rm SBP}\cdot {\bf E}=0$ yields the significantly more accurate solution for the capacitor denoted by the red arrows in the right panel of \cref{fig:CapacitorExample}.
\begin{figure} [h]
\includegraphics[scale=0.3,clip=true, trim= 0 0 0 0 ]{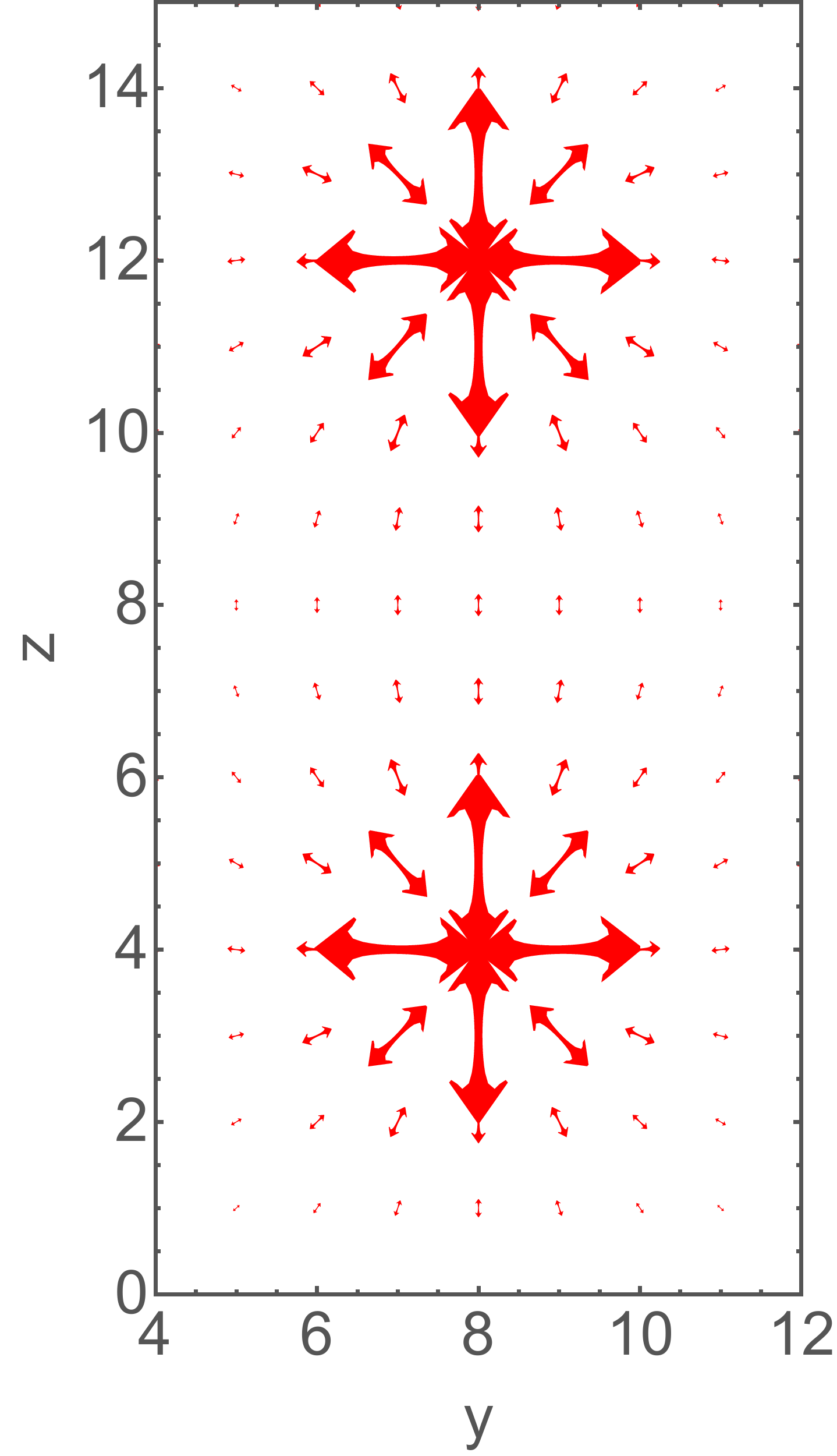}\includegraphics[scale=0.3,clip=true, trim= 0 0 0 0 ]{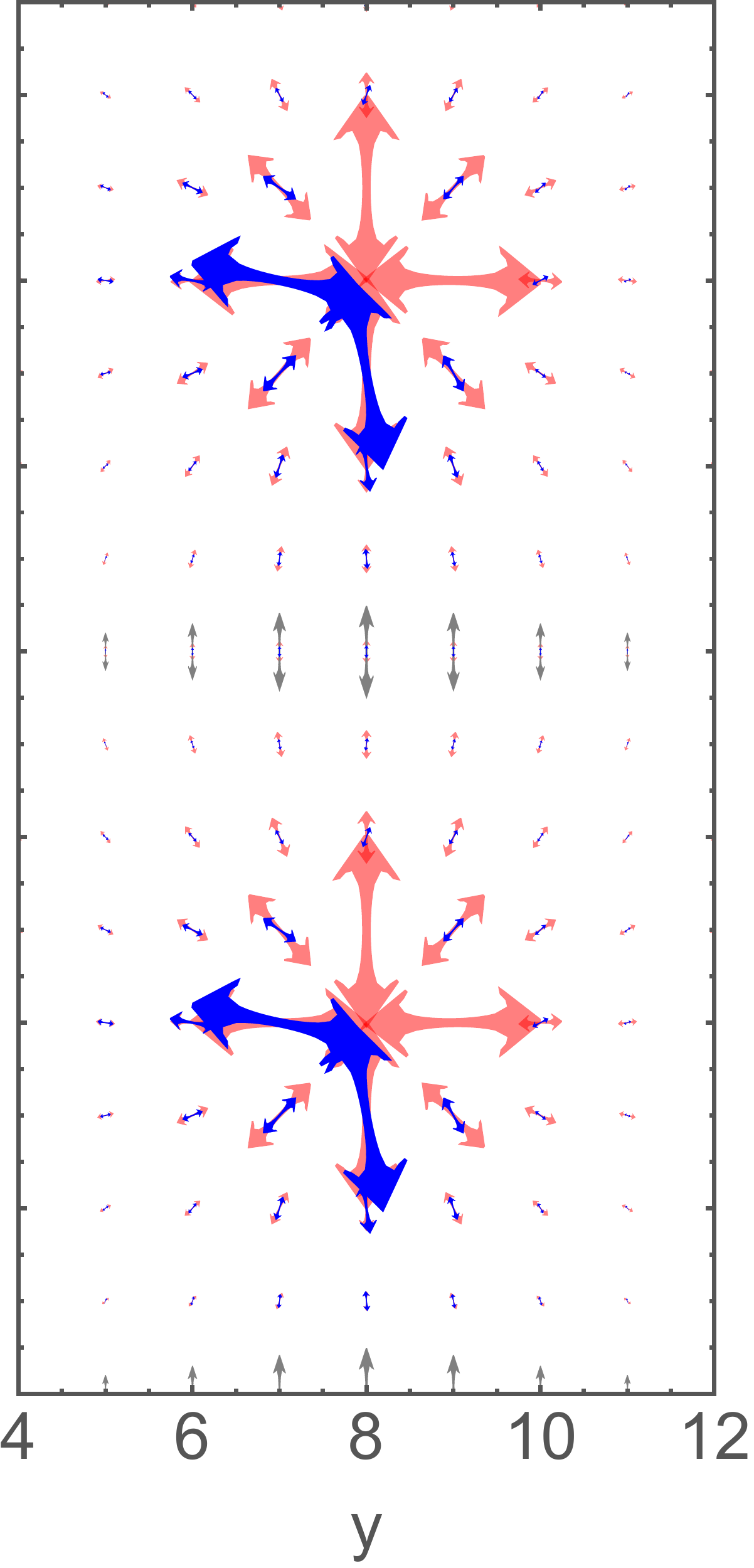}\includegraphics[scale=0.3,clip=true, trim= 0 0 6.3cm 0 ]{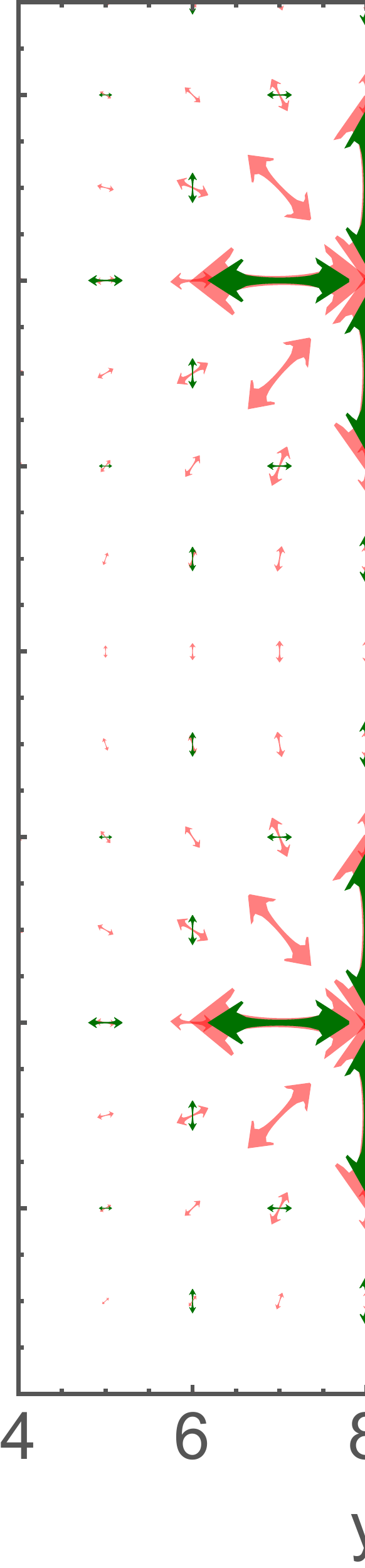}\includegraphics[scale=0.3,clip=true, trim= 5.9cm 0 0 0 ]{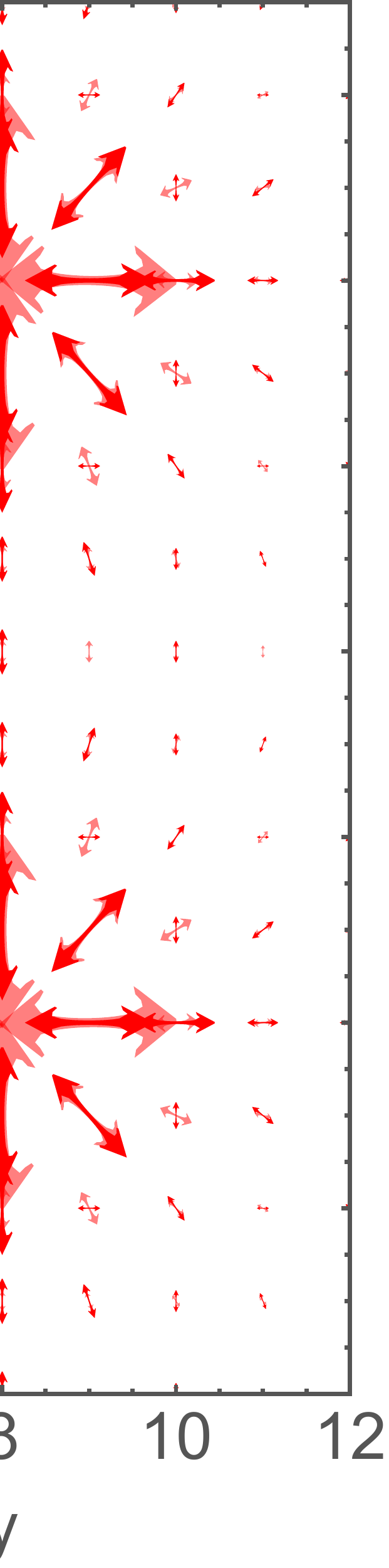}
\caption{(left) Force field lines of the stress tensor between an Abelian unit charge and anti-charge. (center) The EVs obtained from a backward FD solution of Gauss' law. Gray arrows denote the only vectors truly in the yz plane, blue arrows the yz component of those predominantly in the zy plane. Analytic solution is shown in light red (right) solution from the central FD with point sources (green arrows) and the conserving solution with distributed sources (red arrows).} \label{fig:FSTQQExample}
\end{figure}

(B) The second challenge pertains to the fact the naive forward FD approximation of \cref{eq:Waction} introduces significant artifacts (in particular anisotropy) in the energy momentum tensor (EMT). This issue manifests itself already in simple cases, such as in the presence of static charges. The concept of an electric field becomes ambiguous in non-Abelian gauge theory, where one must instead resort to computing the field stress tensor $\Theta^{ij}$, i.e the spatial components of the gauge independent EMT $T^{\alpha\beta}=\frac{1}{4\pi}\big( g^{\alpha\mu}F_{\mu\lambda}F^{\lambda\beta}+\frac{1}{4}g^{\alpha\beta}F_{\mu\lambda}F^{\mu\lambda}\big)$ \cite{Yamamoto:2014vda}. The stress tensor yields the force density ${\bf f}={\bf \nabla}\cdot{\bf \Theta} + \partial {\bf S}/\partial t$ acting on a charge after differentiation and ${\bf S}$ denotes the Poynting vector, which is irrelevant in the static case \cite{Jackson:1998nia}. The force field lines are computed via diagonalizing the stress tensor in Minkowski time. It features two negative and one positive eigenvalue, all of the same magnitude and its eigenvectors (EV) indicate the direction of the forces acting on its faces. Placing a charge at $({\bf x}_0)$ and anti-charge at $({\bf x}_1)$, centered in x and y and separated by a distance in z, the true stress tensor in the centered yz plane, due to symmetry, exhibits two eigenvectors in that plane and one perpendicular to it. The EV associated with the positive eigenvalue indicates the direction of the force lines, as shown in the analytic solution (red arrows) in the left panel of \cref{fig:FSTQQExample}

If instead we compute the stress tensor from the solution of the discrete Gauss law ${\bf \Delta}^{\rm B}\cdot {\bf E}=\frac{1}{a_s^3}\big[ \delta_{{\bf x}{\bf x}_0} - \delta_{{\bf x}{\bf x}_1}\big]$ we find that in the yz plane only a few of its EV (gray arrows in the center panel of \cref{fig:FSTQQExample}) are truly parallel to it. If one plots the vectors with more than 50\% of their magnitude in that plane, one obtains the blue arrows. Compared to the true solution given as light red arrows, a clear asymmetry is visible in the forward FD solution, i.e. the force field is not accurately reproduced. Solving Gauss' law with the central FD discretization we find the symmetric green arrows in the right panel, which however exhibit an artificial pattern of force lines oriented 90 degreed to each other. 

The reason for the failure of the naive central FD approximation lies in its inability to reproduce the integral form of Gauss' law, a fact well known in computational electrodynamics \cite{taflove_computational_2005,bruce_langdon_enforcing_1992}. Indeed the third equality in  $Q = \int dV\, q = \int dV\, ({\bf \nabla} {\bf E} )= \int_{\partial V} d{\bf A} \cdot {\bf E}$ does not hold in general for the central FD. As is standard in modern computational electrodynamics, a genuinely conserving finite volume discretization \cite{leveque_2002}, i.e. one that obeys both the differential and integral Gauss law can be constructed by starting from discretizing the Gauss law with the midpoint rule
\begin{align}
\nonumber\int_{x_{i-1/2}}^{x_{i+1/2}}dx\int_{y_{i-1/2}}^{y_{i+1/2}}dy&\int_{z_{i-1/2}}^{z_{i+1/2}}dz\big( \frac{d E_x}{dx}+\frac{d E_y}{dy}+\frac{d E_z}{dz}\big)\\ &= \int d^3x \delta^{(3)}({\bf x}-{\bf x}_0). \label{eq:FiniteVolDiscr}
\end{align} 
The resulting equations on the LHS can be brought into the form of a central finite difference by adding multiple shifted versions of \cref{eq:FiniteVolDiscr}. This leads us to
\begin{align}
\sum_i \Delta_i^{\rm C} E_i({\bf x})&=\frac{1}{8a^3} \Big[ \sum_i\big( \delta_{{\bf x}+a\hat{\bf i},{\bf x}_0} +\delta_{{\bf x}-a\hat{\bf i},{\bf x}_0}\big)-2\delta_{{\bf x},{\bf x}_0}\Big]. \label{eq:ConsGL}
\end{align}
We see that in order for the central FD approximation to implement the integral Gauss law we must employ a spatially distributed source (c.f. smearing techniques in conventional lattice QCD). \Cref{eq:ConsGL} amounts to a finite volume discretization of Gauss' law. Solving the above yields the symmetric and significantly more accurate solution shown as red arrows on the right of \cref{fig:FSTQQExample}.

We conclude from the Abelian Gauss law that a proper lattice gauge theory for finite systems must both implement the summation by parts property and be formulated with distributed sources. In the remainder of this letter I construct a lattice action that implements these prerequisites.

\section{Lattice gauge theory for finite systems}

\begin{figure} [h]
\includegraphics[scale=0.5,clip=true, trim= 0 0 0 0 ]{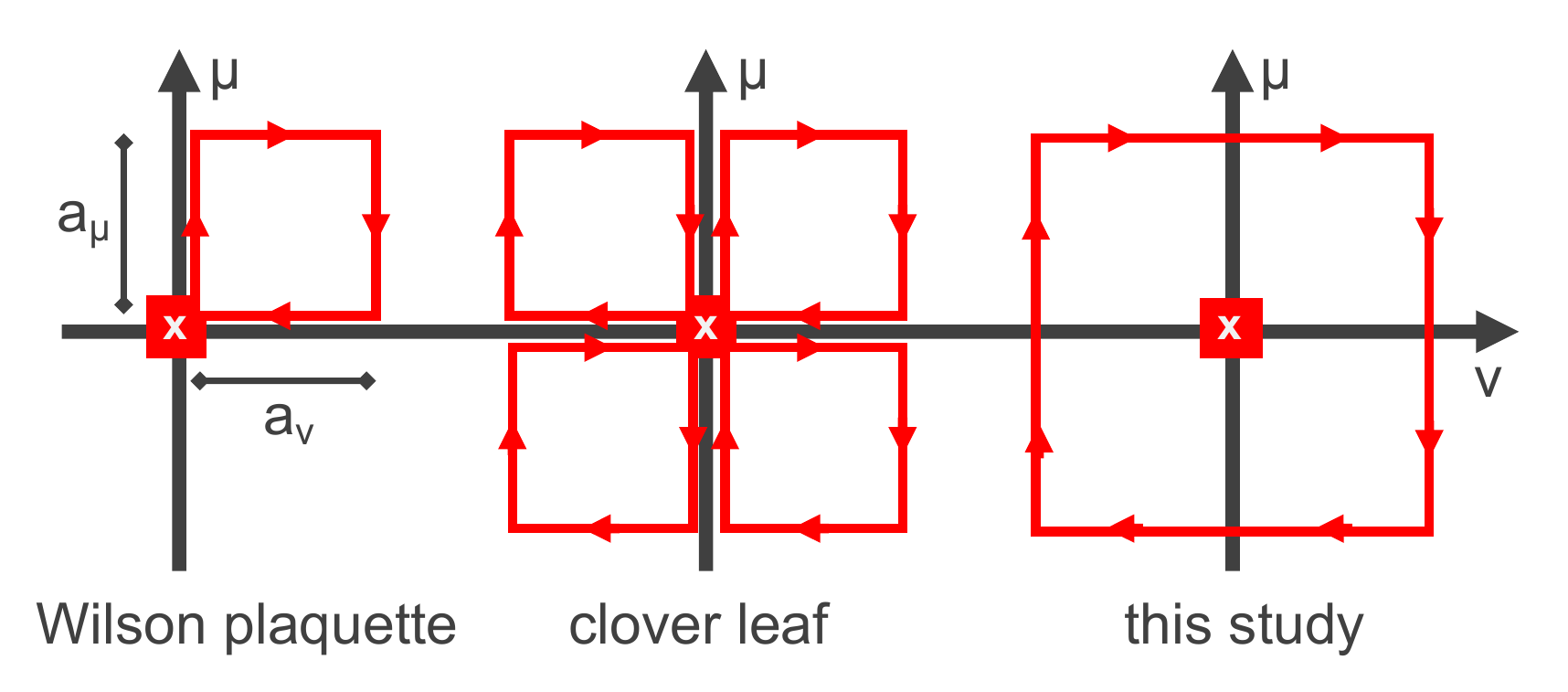}
\caption{Plaquette actions: (left) Wilson (center) clover-leaf (right) stand-alone centered $2\times 2$ plaquette.} \label{fig:DiffAction}
\end{figure}

In the preceding section we used Gauss' law as guide towards an appropriate description for finite systems and discovered that it requires implementing the summation by parts property (A) and the presence of distributed sources (B). To construct such an SBP discretization for non-Abelian gauge theory we need to find the central FD counterpart to \cref{eq:Waction}, operating at global order ${\cal O}(a^2)$ in the lattice spacing. We thus need links of the corresponding order $\bar U_{\mu,x}={\rm exp}\big[igA_{\mu,x+\frac{1}{2}a\hat\mu}\big]={\rm exp}\big[ig\frac{1}{2}\big(A_{\mu,x}+A_{\mu,x+a\hat\mu}\big)\big]+{\cal O}(a^2)$. The central FD requires access to gauge fields in the forward and backward direction in each dimension. The product of eight links (see \cref{fig:DiffAction}) centered around the current spacetime point x includes exactly these contributions 
\begin{align}
\nonumber P^{2\times2}_{\mu\nu,x}=& \bar U_{\mu,x-a\hat\mu-a\hat\nu}\bar U_{\mu,x-a\hat\nu}\bar U_{\nu,x+a\hat\mu-a\hat\nu}\bar U_{\nu,x+a\hat\mu}\times \\
\nonumber &\bar U^\dagger_{\mu,x+a\hat\nu}\bar U^\dagger_{\mu,x-a\hat\mu+a\hat\nu}\bar U^\dagger_{\nu,x-a\hat\mu}\bar U^\dagger_{\nu,x-a\hat\mu-a\hat\nu}\\
\nonumber =&{\rm exp}\big[ 4ig a_\mu a_\nu \bar F_{\mu\nu,x}\big] + {\cal O}(a^3)\\
\bar F_{\mu\nu,x}= &{\bf \Delta}^{\rm C}_\mu A_{\nu,x} - {\bf \Delta}^{\rm C}_\nu A_{\mu,x}+i[A_{\mu,x},A_{\nu,x}]
\end{align}
A gauge invariant action with the correct classical continuum limit may now be constructed from $P^{2\times2}_{\mu\nu,x}$ in the interior of the finite volume V
\begin{align}
 S^{2\times 2}=\sum_{x\notin \partial V} a_ta_s^3\Big[ \frac{2}{16 a_t^2a_s^2}\sum_i{\rm ReTr}\big[ 1-P^{2\times2}_{0i,x}\big]
 - \frac{1}{16 a_s^4}\sum_{ij} {\rm ReTr}\big[ 1-P^{2 \times2}_{ij,x}\big] \Big]
\end{align}
Lattice discretization errors start at one higher power of the lattice spacing as in the Wilson action, making $S^{2\times 2}$ automatically ${\cal O}(a)$ improved. Note the difference to the Symanzik improved actions of Refs. \cite{GarciaPerez:1993lic,Beinlich:1995ik,deForcrand:1997esx,Langfeld:2004tf}, which contain an admixture of asymmetric Wilson plaquettes in the interior and thus do not correspond to a SBP compatible central finite difference.

Next we combine the central FD of the interior with the forward and backward FD on the boundaries, which is achieved by using the forward and backward Wilson plaquettes there in addition to $S^{2\times 2}$, leading finally to
\begin{align}
\tilde F_{\mu\nu,x}= &{\bf \Delta}^{\rm SBP}_\mu A_{\nu,x} - {\bf \Delta}^{\rm SBP}_\nu A_{\mu,x}+i[A_{\mu,x},A_{\nu,x}].
\end{align}
This concludes the solution to challenge (A) for non-Abelian gauge theory. Let me note that simply replacing the Wilson plaquette by the cloverleaf sum does not do the trick: computing $\partial S_{\rm CL}/\partial A_0^a$ leads to the same backward FD scheme for the Gauss law that one obtains from the naive plaquette action itself. $P^{2\times2}$ as product of four shifted $P^{1\times1}$'s adds the exponents, not the exponentials.

Let's now turn to challenge (B). In contrast to computational electrodynamics, which is formulated on the level of the equations of motion, we construct an action based formulation, applicable to the quantum realm. Thus one cannot simply introduce spatially distributed dynamical sources by hand but needs to prescribe how such a term arises from coupling to matter fields. Let me demonstrate the strategy for static matter fields, in which only the covariant derivative in time direction plays a role \cite{brambilla_effective-field_2005}. For static fermions, we may decouple their four-spinor into two Pauli spinors $\Phi=(\psi,\chi)$. Let us focus on the action for $\psi$, as that of $\chi$ follows from complex conjugation: $S_\psi=a_ta_s^3\sum_x i \psi_x^\dagger D_0 \psi_x$. In order to obtain the shift pattern of \cref{eq:ConsGL} required for a conserving discretization, we will have to modify the covariant derivative $D_0$ from its standard form. Taking inspiration from a recently developed FD operator for the simulation of open quantum system dynamics \cite{alund_trace_2021} we may amend the standard temporal derivative on an isotropic lattice by an expression that extends it into one of the spatial directions $j$ as $\Delta^{\rm RN-SBP}_{t(j)}\phi(x)=(\phi(x+a\hat0-a\hat j)-\phi(x-a\hat0-a\hat j)-\phi(x-a\hat0+a\hat j)+\phi(x+a\hat0+a\hat j) )/4a$. One symmetric and gauge covariant formulation of the fermion action thus reads
\begin{align}
S_\psi&=a_ta_s^3\frac{i}{3} \sum_{x,i}  \psi_x^\dagger \bar D_{0(i)} \psi_x= a_ta_s^3\frac{i}{3} \sum_{x,i}  \psi_x^\dagger\frac{1}{8a}\Big[ \\
\nonumber&\big( U_{0,x} U^\dagger_{j,x+a\hat 0-a\hat j}+ U^\dagger_{j,x-a\hat j}U_{0,x-a\hat j}\big)\psi_{x+a\hat 0-a\hat j}\\
\nonumber -&\big( U^\dagger_{j,x-a\hat j}U^\dagger_{0,x-a\hat 0-a\hat j}+ U^\dagger_{0,x-a\hat 0}U^\dagger_{jx-a\hat 0-a\hat j}\big) \psi_{x-a\hat 0-a\hat j}\\
\nonumber -&\big(U^\dagger_{0,x-a\hat 0}U_{j,x-a\hat 0}+U_{j,x}U^\dagger_{0,x-a\hat0+a\hat j}\big) \psi_{x-a\hat0+a\hat j}\\
\nonumber +&(U_{j,x}U_{0,x+a\hat j}+U_{0,x}U_{j,x+a\hat 0}\big)\psi_{x+a\hat 0+a\hat j} \Big]
\end{align}
In the construction of the classical equation of motion in the presence of fermions as outlined in Ref.~\cite{kasper_fermion_2014}, the charge density for $\psi$ is obtained from the expectation value of the fermion two-point function $q_x=\frac{g}{2}{\rm Tr}\big[ \langle [\psi^\dagger_x,\psi_x]\rangle$. Expressing the time derivative as shown above will  introduce symmetric forward and backward shifts in this term, which can be combined into the conserving discretization of \cref{eq:ConsGL}. For truly static charges on the other hand, the spatial distribution of source terms in \Cref{eq:ConsGL} can be implemented directly through a modification of the source terms within the Wilson loop. 


\section{Summary \& Conclusion}
I have argued and presented numerical evidence in the Abelian theory that in order to capture the physics of non-trivial boundary conditions, the naive forward finite difference discretization underlying the Wilson action must be replaced with one that respects the summation by parts property. In addition, even in the presence of trivial periodic boundary conditions, the asymmetric Wilson action, as well as the clover leaf, both lead to a backward finite difference Gauss law, which as I have shown imprints asymmetries into the force field lines encoded in the energy momentum tensor. The standard Symanzik improvement, while reducing the magnitude of the discretization error, does not alleviate the asymmetry of the underlying discretization. The naive central finite difference discretization however does not suffice for a faithful reproduction of field line geometry and instead I have argued that a finite volume formulation using distributed sources is called for. To address both issues, I thus proposed a novel central finite difference discretization of the gauge action based on a stand-alone 2-by-2 plaquette centered around the point of evaluation. Combined with forward and backward plaquettes on the boundary, a summation by parts discretized field strength tensor ensues. I furthermore propose to implement the distributed source structure by a modification of the derivative terms in the fermion action, which shifts the derivatives into adjacent planes.

$S^{2\times 2}$ is easily deployed in existing simulation codes for Euclidean time. For its use in real-time applications, which are most conveniently formulated in a Hamiltonian language, we may straight forwardly derive a continuous time variant. While the leap-frog type equations of motion from the naive Wilson action automatically preserve the differential Gauss-law up to rounding errors, the spatially extended nature of $S^{2\times 2}$ leads to a non-trivial consistency condition in the Dirac-Bergman sense, which has to be accommodated by a modified RATTLE algorithm \cite{andersen_rattle_1983}, which is work in progress. 

The novel action (interior $2\times2$ and boundary $1\times1$ plaquettes with distributed sources) presented in this letter for the first time allows the powerful machinery of lattice gauge theory to be consistently applied to systems with non-trivial boundary conditions, opening up new venues of investigation of non-perturbative phenomena in lattice QCD and cavity QED. At the same time it constitutes a genuine second order symmetric discretization of lattice gauge theory, offering the mathematical community a new impulse in their ongoing quest for a rigorous formulation of higher order discrete gauge invariance \cite{christiansen_finite_2020,schubel2018discretization,christiansen_second_2015}.

\section*{Acknowledgements}
The author thanks Sigbj\o rn Hervik, Jan Nordstr\"om, Jan Pawlowski and Anders Tranberg for valuable discussion and acknowledges funding from the Research Council of Norway under the FRIPRO Young Research Talent grant 286883.


\FloatBarrier

\begin{backmatter}

\section*{Competing interests}
  The author declares that he has no competing interests.

\section*{Author's contributions}
    \begin{itemize}
        \item A. Rothkopf: conceptualization, implementation, writing
    \end{itemize}


\bibliographystyle{stavanger-mathphys}


\bibliography{LGTforFiniteSystems}


\begin{thebibliography}{37}
\ifx \bisbn   \undefined \def \bisbn  #1{ISBN #1}\fi
\ifx \binits  \undefined \def \binits#1{#1}\fi
\ifx \bauthor  \undefined \def \bauthor#1{#1}\fi
\ifx \batitle  \undefined \def \batitle#1{#1}\fi
\ifx \bjtitle  \undefined \def \bjtitle#1{#1}\fi
\ifx \bvolume  \undefined \def \bvolume#1{\textbf{#1}}\fi
\ifx \byear  \undefined \def \byear#1{#1}\fi
\ifx \bissue  \undefined \def \bissue#1{#1}\fi
\ifx \bfpage  \undefined \def \bfpage#1{#1}\fi
\ifx \blpage  \undefined \def \blpage #1{#1}\fi
\ifx \burl  \undefined \def \burl#1{\textsf{#1}}\fi
\ifx \doiurl  \undefined \def \doiurl#1{\textsf{#1}}\fi
\ifx \betal  \undefined \def \betal{\textit{et al.}}\fi
\ifx \binstitute  \undefined \def \binstitute#1{#1}\fi
\ifx \binstitutionaled  \undefined \def \binstitutionaled#1{#1}\fi
\ifx \bctitle  \undefined \def \bctitle#1{#1}\fi
\ifx \beditor  \undefined \def \beditor#1{#1}\fi
\ifx \bpublisher  \undefined \def \bpublisher#1{#1}\fi
\ifx \bbtitle  \undefined \def \bbtitle#1{#1}\fi
\ifx \bedition  \undefined \def \bedition#1{#1}\fi
\ifx \bseriesno  \undefined \def \bseriesno#1{#1}\fi
\ifx \blocation  \undefined \def \blocation#1{#1}\fi
\ifx \bsertitle  \undefined \def \bsertitle#1{#1}\fi
\ifx \bsnm \undefined \def \bsnm#1{#1}\fi
\ifx \bsuffix \undefined \def \bsuffix#1{#1}\fi
\ifx \bparticle \undefined \def \bparticle#1{#1}\fi
\ifx \barticle \undefined \def \barticle#1{#1}\fi
\ifx \bconfdate \undefined \def \bconfdate #1{#1}\fi
\ifx \botherref \undefined \def \botherref #1{#1}\fi
\ifx \url \undefined \def \url#1{\textsf{#1}}\fi
\ifx \bchapter \undefined \def \bchapter#1{#1}\fi
\ifx \bbook \undefined \def \bbook#1{#1}\fi
\ifx \bcomment \undefined \def \bcomment#1{#1}\fi
\ifx \oauthor \undefined \def \oauthor#1{#1}\fi
\ifx \citeauthoryear \undefined \def \citeauthoryear#1{#1}\fi
\ifx \endbibitem  \undefined \def \endbibitem {}\fi
\ifx \bconflocation  \undefined \def \bconflocation#1{#1}\fi
\ifx \arxivurl  \undefined \def \arxivurl#1{\textsf{#1}}\fi
\csname PreBibitemsHook\endcsname

\bibitem{detmold_topical_2019}
\begin{barticle}
\bauthor{\bsnm{Detmold}, \binits{W.}},
\bauthor{\bsnm{Kronfeld}, \binits{A.}},
\bauthor{\bsnm{Mei\ss~ner}, \binits{U.-G.}}:
\batitle{Topical issue on opportunities for lattice gauge theory in the era of
  exascale computing}.
\bjtitle{The European Physical Journal A}
\bvolume{55}(\bissue{11}),
\bfpage{192}
(\byear{2019}).
doi:\doiurl{10.1140/epja/i2019-12942-8}.
Accessed 2021-02-14
\end{barticle}
\endbibitem

\bibitem{gpd_lat_2020}
\begin{barticle}
\bauthor{\bsnm{{[Extended Twisted Mass Collaboration]}}},
\bauthor{\bsnm{Alexandrou}, \binits{C.}}, \betal:
\batitle{Unpolarized and {Helicity} {Generalized} {Parton} {Distributions} of
  the {Proton} within {Lattice} {QCD}}.
\bjtitle{Physical Review Letters}
\bvolume{125}(\bissue{26}),
\bfpage{262001}
(\byear{2020}).
doi:\doiurl{10.1103/PhysRevLett.125.262001}.
Accessed 2021-02-14
\end{barticle}
\endbibitem

\bibitem{borsanyi_ab_2015}
\begin{barticle}
\bauthor{\bsnm{Borsanyi}, \binits{S.}}, \betal:
\batitle{Ab initio calculation of the neutron-proton mass difference}.
\bjtitle{Science}
\bvolume{347}(\bissue{6229}),
\bfpage{1452}--\blpage{1455}
(\byear{2015}).
doi:\doiurl{10.1126/science.1257050}
\end{barticle}
\endbibitem

\bibitem{briceno_scattering_2018}
\begin{barticle}
\bauthor{\bsnm{Brice\~no}, \binits{R.A.}},
\bauthor{\bsnm{Dudek}, \binits{J.J.}},
\bauthor{\bsnm{Young}, \binits{R.D.}}:
\batitle{Scattering processes and resonances from lattice {QCD}}.
\bjtitle{Reviews of Modern Physics}
\bvolume{90}(\bissue{2}),
\bfpage{025001}
(\byear{2018}).
doi:\doiurl{10.1103/RevModPhys.90.025001}.
Accessed 2021-02-14
\end{barticle}
\endbibitem

\bibitem{bazavov_hot-dense_2019}
\begin{barticle}
\bauthor{\bsnm{Bazavov}, \binits{A.}},
\bauthor{\bsnm{Karsch}, \binits{F.}},
\bauthor{\bsnm{Mukherjee}, \binits{S.}},
\bauthor{\bsnm{Petreczky}, \binits{P.}},
\bauthor{\bsnm{{USQCD Collaboration}}}:
\batitle{Hot-dense lattice qcd}.
\bjtitle{The European Physical Journal A}
\bvolume{55}(\bissue{11}),
\bfpage{194}
(\byear{2019}).
doi:\doiurl{10.1140/epja/i2019-12922-0}.
Accessed 2021-02-14
\end{barticle}
\endbibitem

\bibitem{axion_lat_2016}
\begin{barticle}
\bauthor{\bsnm{Borsanyi}, \binits{S.}}, \betal:
\batitle{Calculation of the axion mass based on high-temperature lattice
  quantum chromodynamics}.
\bjtitle{Nature}
\bvolume{539}(\bissue{7627}),
\bfpage{69}--\blpage{71}
(\byear{2016}).
doi:\doiurl{10.1038/nature20115}.
Accessed 2021-02-14
\end{barticle}
\endbibitem

\bibitem{cQED1_2019}
\begin{barticle}
\bauthor{\bsnm{Kockum}, \binits{F.}}, \betal:
\batitle{Ultrastrong coupling between light and matter}.
\bjtitle{Nature Reviews Physics}
\bvolume{1}(\bissue{1}),
\bfpage{19}--\blpage{40}
(\byear{2019}).
doi:\doiurl{10.1038/s42254-018-0006-2}.
Accessed 2021-02-14
\end{barticle}
\endbibitem

\bibitem{cQED2_2019}
\begin{barticle}
\bauthor{\bsnm{Forn-D\'iaz}, \binits{P.}},
\bauthor{\bsnm{Lamata}, \binits{L.}},
\bauthor{\bsnm{Rico}, \binits{E.}},
\bauthor{\bsnm{Kono}, \binits{J.}},
\bauthor{\bsnm{Solano}, \binits{E.}}:
\batitle{Ultrastrong coupling regimes of light-matter interaction}.
\bjtitle{Reviews of Modern Physics}
\bvolume{91}(\bissue{2}),
\bfpage{025005}
(\byear{2019}).
doi:\doiurl{10.1103/RevModPhys.91.025005}.
Accessed 2021-02-14
\end{barticle}
\endbibitem

\bibitem{jaako_quantum_2020}
\begin{barticle}
\bauthor{\bsnm{Jaako}, \binits{T.}},
\bauthor{\bsnm{Garcia‐Ripoll}, \binits{J.J.}},
\bauthor{\bsnm{Rabl}, \binits{P.}}:
\batitle{Quantum {Simulation} of {Non}-{Perturbative} {Cavity} {QED} with
  {Trapped} {Ions}}.
\bjtitle{Advanced Quantum Technologies}
\bvolume{3}(\bissue{4}),
\bfpage{1900125}
(\byear{2020}).
doi:\doiurl{10.1002/qute.201900125}.
Accessed 2021-02-14
\end{barticle}
\endbibitem

\bibitem{nagle_small_2018}
\begin{barticle}
\bauthor{\bsnm{Nagle}, \binits{J.L.}},
\bauthor{\bsnm{Zajc}, \binits{W.A.}}:
\batitle{Small {System} {Collectivity} in {Relativistic} {Hadronic} and
  {Nuclear} {Collisions}}.
\bjtitle{Annual Review of Nuclear and Particle Science}
\bvolume{68}(\bissue{1}),
\bfpage{211}--\blpage{235}
(\byear{2018}).
doi:\doiurl{10.1146/annurev-nucl-101916-123209}.
Accessed 2021-02-14
\end{barticle}
\endbibitem

\bibitem{mogliacci_geometrically_2020}
\begin{barticle}
\bauthor{\bsnm{Mogliacci}, \binits{S.}},
\bauthor{\bsnm{Kolbe}, \binits{I.}},
\bauthor{\bsnm{Horowitz}, \binits{W.}}:
\batitle{Geometrically confined thermal field theory: {Finite} size corrections
  and phase transitions}.
\bjtitle{Physical Review D}
\bvolume{102}(\bissue{11}),
\bfpage{116017}
(\byear{2020}).
doi:\doiurl{10.1103/PhysRevD.102.116017}.
Accessed 2021-02-14
\end{barticle}
\endbibitem

\bibitem{Panero:2008mg}
\begin{barticle}
\bauthor{\bsnm{Panero}, \binits{M.}}:
\batitle{{Geometric effects in lattice QCD thermodynamics}}.
\bjtitle{PoS}
\bvolume{LATTICE2008},
\bfpage{175}
(\byear{2008}).
doi:\doiurl{10.22323/1.066.0175}
\end{barticle}
\endbibitem

\bibitem{casanova_deep_2010}
\begin{barticle}
\bauthor{\bsnm{Casanova}, \binits{J.}}, \betal:
\batitle{Deep {Strong} {Coupling} {Regime} of the {Jaynes}-{Cummings} {Model}}.
\bjtitle{Physical Review Letters}
\bvolume{105}(\bissue{26}),
\bfpage{263603}
(\byear{2010}).
doi:\doiurl{10.1103/PhysRevLett.105.263603}.
Accessed 2021-02-14
\end{barticle}
\endbibitem

\bibitem{wilson_confinement_1974}
\begin{barticle}
\bauthor{\bsnm{Wilson}, \binits{K.G.}}:
\batitle{Confinement of quarks}.
\bjtitle{Physical Review D}
\bvolume{10}(\bissue{8}),
\bfpage{2445}--\blpage{2459}
(\byear{1974}).
doi:\doiurl{10.1103/PhysRevD.10.2445}.
Accessed 2021-02-14
\end{barticle}
\endbibitem

\bibitem{klassen_anisotropic_1998}
\begin{barticle}
\bauthor{\bsnm{Klassen}, \binits{T.R.}}:
\batitle{The anisotropic {Wilson} gauge action}.
\bjtitle{Nuclear Physics B}
\bvolume{533}(\bissue{1}),
\bfpage{557}--\blpage{575}
(\byear{1998}).
doi:\doiurl{10.1016/S0550-3213(98)00510-0}.
Accessed 2021-02-14
\end{barticle}
\endbibitem

\bibitem{symanzik_continuum_1983}
\begin{barticle}
\bauthor{\bsnm{Symanzik}, \binits{K.}}:
\batitle{Continuum limit and improved action in lattice theories: ({I}).
  {Principles} and $\phi^4$ theory}.
\bjtitle{Nuclear Physics B}
\bvolume{226}(\bissue{1}),
\bfpage{187}--\blpage{204}
(\byear{1983}).
doi:\doiurl{10.1016/0550-3213(83)90468-6}.
Accessed 2021-02-14
\end{barticle}
\endbibitem

\bibitem{GarciaPerez:1993lic}
\begin{barticle}
\bauthor{\bsnm{Garcia~Perez}, \binits{M.}},
\bauthor{\bsnm{Gonzalez-Arroyo}, \binits{A.}},
\bauthor{\bsnm{Snippe}, \binits{J.R.}},
\bauthor{\bparticle{van} \bsnm{Baal}, \binits{P.}}:
\batitle{{Instantons from over - improved cooling}}.
\bjtitle{Nucl. Phys. B}
\bvolume{413},
\bfpage{535}--\blpage{552}
(\byear{1994}).
doi:\doiurl{10.1016/0550-3213(94)90631-9}
\end{barticle}
\endbibitem

\bibitem{Beinlich:1995ik}
\begin{barticle}
\bauthor{\bsnm{Beinlich}, \binits{B.}},
\bauthor{\bsnm{Karsch}, \binits{F.}},
\bauthor{\bsnm{Laermann}, \binits{E.}}:
\batitle{{Improved actions for QCD thermodynamics on the lattice}}.
\bjtitle{Nucl. Phys. B}
\bvolume{462},
\bfpage{415}--\blpage{436}
(\byear{1996}).
doi:\doiurl{10.1016/0550-3213(95)00681-8}
\end{barticle}
\endbibitem

\bibitem{deForcrand:1997esx}
\begin{barticle}
\bauthor{\bparticle{de} \bsnm{Forcrand}, \binits{P.}},
\bauthor{\bsnm{Garcia~Perez}, \binits{M.}},
\bauthor{\bsnm{Stamatescu}, \binits{I.-O.}}:
\batitle{{Topology of the SU(2) vacuum: A Lattice study using improved
  cooling}}.
\bjtitle{Nucl. Phys. B}
\bvolume{499},
\bfpage{409}--\blpage{449}
(\byear{1997}).
doi:\doiurl{10.1016/S0550-3213(97)00275-7}
\end{barticle}
\endbibitem

\bibitem{Snippe:1997ru}
\begin{barticle}
\bauthor{\bsnm{Snippe}, \binits{J.R.}}:
\batitle{{Computation of the one loop Symanzik coefficients for the square
  action}}.
\bjtitle{Nucl. Phys. B}
\bvolume{498},
\bfpage{347}--\blpage{396}
(\byear{1997}).
doi:\doiurl{10.1016/S0550-3213(97)00270-8}
\end{barticle}
\endbibitem

\bibitem{BilsonThompson:2002jk}
\begin{barticle}
\bauthor{\bsnm{Bilson-Thompson}, \binits{S.O.}},
\bauthor{\bsnm{Leinweber}, \binits{D.B.}},
\bauthor{\bsnm{Williams}, \binits{A.G.}}:
\batitle{{Highly improved lattice field strength tensor}}.
\bjtitle{Annals Phys.}
\bvolume{304},
\bfpage{1}--\blpage{21}
(\byear{2003}).
doi:\doiurl{10.1016/S0003-4916(03)00009-5}
\end{barticle}
\endbibitem

\bibitem{Langfeld:2004tf}
\begin{botherref}
\oauthor{\bsnm{Langfeld}, \binits{K.}}:
{A Novel improved action for SU(3) lattice gauge theory}
(2004).
\arxivurl{hep-lat/0403018}
\end{botherref}
\endbibitem

\bibitem{Jansen:2008vs}
\begin{barticle}
\bauthor{\bsnm{Jansen}, \binits{K.}}:
\batitle{{Lattice QCD: A Critical status report}}.
\bjtitle{PoS}
\bvolume{LATTICE2008},
\bfpage{010}
(\byear{2008}).
doi:\doiurl{10.22323/1.066.0010}.
\arxivurl{0810.5634}
\end{barticle}
\endbibitem

\bibitem{svard2014review}
\begin{barticle}
\bauthor{\bsnm{Sv{\"a}rd}, \binits{M.}},
\bauthor{\bsnm{Nordstr{\"o}m}, \binits{J.}}:
\batitle{Review of summation-by-parts schemes for initial--boundary-value
  problems}.
\bjtitle{Journal of Computational Physics}
\bvolume{268},
\bfpage{17}--\blpage{38}
(\byear{2014})
\end{barticle}
\endbibitem

\bibitem{fernandez2014review}
\begin{barticle}
\bauthor{\bsnm{Fern{\'a}ndez}, \binits{D.C.D.R.}},
\bauthor{\bsnm{Hicken}, \binits{J.E.}},
\bauthor{\bsnm{Zingg}, \binits{D.W.}}:
\batitle{Review of summation-by-parts operators with simultaneous approximation
  terms for the numerical solution of partial differential equations}.
\bjtitle{Computers \& Fluids}
\bvolume{95},
\bfpage{171}--\blpage{196}
(\byear{2014})
\end{barticle}
\endbibitem

\bibitem{Yamamoto:2014vda}
\begin{barticle}
\bauthor{\bsnm{Yamamoto}, \binits{A.}}:
\batitle{{Lattice QCD in curved spacetimes}}.
\bjtitle{Phys. Rev. D}
\bvolume{90}(\bissue{5}),
\bfpage{054510}
(\byear{2014}).
doi:\doiurl{10.1103/PhysRevD.90.054510}.
\arxivurl{1405.6665}
\end{barticle}
\endbibitem

\bibitem{Jackson:1998nia}
\begin{bbook}
\bauthor{\bsnm{Jackson}, \binits{J.D.}}:
\bbtitle{{Classical Electrodynamics}}.
\bpublisher{Wiley}, \blocation{???}
(\byear{1998})
\end{bbook}
\endbibitem

\bibitem{taflove_computational_2005}
\begin{bbook}
\bauthor{\bsnm{Taflove}, \binits{A.}},
\bauthor{\bsnm{Hagness}, \binits{S.C.}}:
\bbtitle{Computational {Electrodynamics}: {The} {Finite}-difference
  {Time}-domain {Method}}.
\bsertitle{Artech {House} antennas and propagation library}.
\bpublisher{Artech House}, \blocation{???}
(\byear{2005}).
\burl{https://books.google.no/books?id=n2ViQgAACAAJ}
\end{bbook}
\endbibitem

\bibitem{bruce_langdon_enforcing_1992}
\begin{barticle}
\bauthor{\bsnm{Bruce~Langdon}, \binits{A.}}:
\batitle{On enforcing {Gauss}' law in electromagnetic particle-in-cell codes}.
\bjtitle{Computer Physics Communications}
\bvolume{70}(\bissue{3}),
\bfpage{447}--\blpage{450}
(\byear{1992}).
doi:\doiurl{10.1016/0010-4655(92)90105-8}.
Accessed 2021-02-14
\end{barticle}
\endbibitem

\bibitem{leveque_2002}
\begin{botherref}
\oauthor{\bsnm{LeVeque}, \binits{R.J.}}:
Finite volume methods for hyperbolic problems
(2002).
doi:\doiurl{10.1017/CBO9780511791253}
\end{botherref}
\endbibitem

\bibitem{brambilla_effective-field_2005}
\begin{barticle}
\bauthor{\bsnm{Brambilla}, \binits{N.}},
\bauthor{\bsnm{Pineda}, \binits{A.}},
\bauthor{\bsnm{Soto}, \binits{J.}},
\bauthor{\bsnm{Vairo}, \binits{A.}}:
\batitle{Effective-field theories for heavy quarkonium}.
\bjtitle{Reviews of Modern Physics}
\bvolume{77}(\bissue{4}),
\bfpage{1423}--\blpage{1496}
(\byear{2005}).
doi:\doiurl{10.1103/RevModPhys.77.1423}.
Accessed 2021-02-14
\end{barticle}
\endbibitem

\bibitem{alund_trace_2021}
\begin{barticle}
\bauthor{\bsnm{Ålund}, \binits{O.}}, \betal:
\batitle{Trace preserving quantum dynamics using a novel
  reparametrization-neutral summation-by-parts difference operator}.
\bjtitle{Journal of Computational Physics}
\bvolume{425},
\bfpage{109917}
(\byear{2021}).
doi:\doiurl{10.1016/j.jcp.2020.109917}.
Accessed 2021-02-14
\end{barticle}
\endbibitem

\bibitem{kasper_fermion_2014}
\begin{barticle}
\bauthor{\bsnm{Kasper}, \binits{V.}},
\bauthor{\bsnm{Hebenstreit}, \binits{F.}},
\bauthor{\bsnm{Berges}, \binits{J.}}:
\batitle{Fermion production from real-time lattice gauge theory in the
  classical-statistical regime}.
\bjtitle{Physical Review D}
\bvolume{90}(\bissue{2}),
\bfpage{025016}
(\byear{2014}).
doi:\doiurl{10.1103/PhysRevD.90.025016}.
Accessed 2020-12-06
\end{barticle}
\endbibitem

\bibitem{andersen_rattle_1983}
\begin{barticle}
\bauthor{\bsnm{Andersen}, \binits{H.C.}}:
\batitle{Rattle: {A} “velocity” version of the shake algorithm for
  molecular dynamics calculations}.
\bjtitle{Journal of Computational Physics}
\bvolume{52}(\bissue{1}),
\bfpage{24}--\blpage{34}
(\byear{1983}).
doi:\doiurl{10.1016/0021-9991(83)90014-1}.
Accessed 2021-02-14
\end{barticle}
\endbibitem

\bibitem{christiansen_finite_2020}
\begin{botherref}
\oauthor{\bsnm{Christiansen}, \binits{S.H.}},
\oauthor{\bsnm{Hu}, \binits{K.}}:
Finite {Element} {Systems} for vector bundles : elasticity and curvature.
arXiv:1906.09128 [cs, math]
(2020).
Accessed 2021-02-12
\end{botherref}
\endbibitem

\bibitem{schubel2018discretization}
\begin{botherref}
\oauthor{\bsnm{Schubel}, \binits{M.D.}}:
Discretization of differential geometry for computational gauge theory.
PhD thesis,
University of Illinois at Urbana-Champaign
(2018)
\end{botherref}
\endbibitem

\bibitem{christiansen_second_2015}
\begin{botherref}
\oauthor{\bsnm{Christiansen}, \binits{S.H.}},
\oauthor{\bsnm{Halvorsen}, \binits{T.G.}}:
Second order gauge invariant discretizations to the
  {Schr}{\textbackslash}"odinger and {Pauli} equations.
arXiv:1505.08040 [math]
(2015).
Accessed 2021-02-12
\end{botherref}
\endbibitem

\end{thebibliography}

\newcommand{\BMCxmlcomment}[1]{}

\BMCxmlcomment{

<refgrp>

<bibl id="B1">
  <title><p>Topical Issue on Opportunities for Lattice Gauge Theory in the Era
  of Exascale Computing</p></title>
  <aug>
    <au><snm>Detmold</snm><fnm>W</fnm></au>
    <au><snm>Kronfeld</snm><fnm>A</fnm></au>
    <au><snm>Mei\ss ner</snm><fnm>UG</fnm></au>
  </aug>
  <source>The European Physical Journal A</source>
  <pubdate>2019</pubdate>
  <volume>55</volume>
  <issue>11</issue>
  <fpage>192</fpage>
  <url>https://doi.org/10.1140/epja/i2019-12942-8</url>
</bibl>

<bibl id="B2">
  <title><p>Unpolarized and {Helicity} {Generalized} {Parton} {Distributions}
  of the {Proton} within {Lattice} {QCD}</p></title>
  <aug>
    <au><cnm>{[Extended Twisted Mass Collaboration]}</cnm></au>
    <au><snm>Alexandrou</snm><fnm>C</fnm></au>
    <au><cnm>others</cnm></au>
  </aug>
  <source>Physical Review Letters</source>
  <pubdate>2020</pubdate>
  <volume>125</volume>
  <issue>26</issue>
  <fpage>262001</fpage>
  <url>https://link.aps.org/doi/10.1103/PhysRevLett.125.262001</url>
</bibl>

<bibl id="B3">
  <title><p>Ab initio calculation of the neutron-proton mass
  difference</p></title>
  <aug>
    <au><snm>Borsanyi</snm><fnm>S.</fnm></au>
    <au><cnm>others</cnm></au>
  </aug>
  <source>Science</source>
  <pubdate>2015</pubdate>
  <volume>347</volume>
  <issue>6229</issue>
  <fpage>1452</fpage>
  <lpage>-1455</lpage>
  <url>http://dx.doi.org/10.1126/science.1257050</url>
</bibl>

<bibl id="B4">
  <title><p>Scattering processes and resonances from lattice {QCD}</p></title>
  <aug>
    <au><snm>Brice\ no</snm><fnm>RA</fnm></au>
    <au><snm>Dudek</snm><fnm>JJ</fnm></au>
    <au><snm>Young</snm><fnm>RD</fnm></au>
  </aug>
  <source>Reviews of Modern Physics</source>
  <pubdate>2018</pubdate>
  <volume>90</volume>
  <issue>2</issue>
  <fpage>025001</fpage>
  <url>https://link.aps.org/doi/10.1103/RevModPhys.90.025001</url>
</bibl>

<bibl id="B5">
  <title><p>Hot-dense Lattice QCD</p></title>
  <aug>
    <au><snm>Bazavov</snm><fnm>A</fnm></au>
    <au><snm>Karsch</snm><fnm>F</fnm></au>
    <au><snm>Mukherjee</snm><fnm>S</fnm></au>
    <au><snm>Petreczky</snm><fnm>P</fnm></au>
    <au><cnm>{USQCD Collaboration}</cnm></au>
  </aug>
  <source>The European Physical Journal A</source>
  <pubdate>2019</pubdate>
  <volume>55</volume>
  <issue>11</issue>
  <fpage>194</fpage>
  <url>https://doi.org/10.1140/epja/i2019-12922-0</url>
</bibl>

<bibl id="B6">
  <title><p>Calculation of the axion mass based on high-temperature lattice
  quantum chromodynamics</p></title>
  <aug>
    <au><snm>Borsanyi</snm><fnm>S.</fnm></au>
    <au><cnm>others</cnm></au>
  </aug>
  <source>Nature</source>
  <pubdate>2016</pubdate>
  <volume>539</volume>
  <issue>7627</issue>
  <fpage>69</fpage>
  <lpage>-71</lpage>
  <url>https://www.nature.com/articles/nature20115</url>
</bibl>

<bibl id="B7">
  <title><p>Ultrastrong coupling between light and matter</p></title>
  <aug>
    <au><snm>Kockum</snm><fnm>F</fnm></au>
    <au><cnm>others</cnm></au>
  </aug>
  <source>Nature Reviews Physics</source>
  <pubdate>2019</pubdate>
  <volume>1</volume>
  <issue>1</issue>
  <fpage>19</fpage>
  <lpage>-40</lpage>
  <url>https://www.nature.com/articles/s42254-018-0006-2</url>
</bibl>

<bibl id="B8">
  <title><p>Ultrastrong coupling regimes of light-matter
  interaction</p></title>
  <aug>
    <au><snm>Forn D\'iaz</snm><fnm>P.</fnm></au>
    <au><snm>Lamata</snm><fnm>L.</fnm></au>
    <au><snm>Rico</snm><fnm>E.</fnm></au>
    <au><snm>Kono</snm><fnm>J.</fnm></au>
    <au><snm>Solano</snm><fnm>E.</fnm></au>
  </aug>
  <source>Reviews of Modern Physics</source>
  <pubdate>2019</pubdate>
  <volume>91</volume>
  <issue>2</issue>
  <fpage>025005</fpage>
  <url>https://link.aps.org/doi/10.1103/RevModPhys.91.025005</url>
</bibl>

<bibl id="B9">
  <title><p>Quantum {Simulation} of {Non}-{Perturbative} {Cavity} {QED} with
  {Trapped} {Ions}</p></title>
  <aug>
    <au><snm>Jaako</snm><fnm>T</fnm></au>
    <au><snm>Garcia‐Ripoll</snm><fnm>JJ</fnm></au>
    <au><snm>Rabl</snm><fnm>P</fnm></au>
  </aug>
  <source>Advanced Quantum Technologies</source>
  <pubdate>2020</pubdate>
  <volume>3</volume>
  <issue>4</issue>
  <fpage>1900125</fpage>
  <url>https://onlinelibrary.wiley.com/doi/abs/10.1002/qute.201900125</url>
</bibl>

<bibl id="B10">
  <title><p>Small {System} {Collectivity} in {Relativistic} {Hadronic} and
  {Nuclear} {Collisions}</p></title>
  <aug>
    <au><snm>Nagle</snm><fnm>JL</fnm></au>
    <au><snm>Zajc</snm><fnm>WA</fnm></au>
  </aug>
  <source>Annual Review of Nuclear and Particle Science</source>
  <pubdate>2018</pubdate>
  <volume>68</volume>
  <issue>1</issue>
  <fpage>211</fpage>
  <lpage>-235</lpage>
  <url>https://www.annualreviews.org/doi/10.1146/annurev-nucl-101916-123209</url>
</bibl>

<bibl id="B11">
  <title><p>Geometrically confined thermal field theory: {Finite} size
  corrections and phase transitions</p></title>
  <aug>
    <au><snm>Mogliacci</snm><fnm>S</fnm></au>
    <au><snm>Kolbe</snm><fnm>I</fnm></au>
    <au><snm>Horowitz</snm><fnm>W</fnm></au>
  </aug>
  <source>Physical Review D</source>
  <pubdate>2020</pubdate>
  <volume>102</volume>
  <issue>11</issue>
  <fpage>116017</fpage>
  <url>https://link.aps.org/doi/10.1103/PhysRevD.102.116017</url>
</bibl>

<bibl id="B12">
  <title><p>{Geometric effects in lattice QCD thermodynamics}</p></title>
  <aug>
    <au><snm>Panero</snm><fnm>M</fnm></au>
  </aug>
  <source>PoS</source>
  <editor>Aubin, Christopher and Cohen, Saul and Dawson, Chris and Dudek, Jozef
  and Edwards, Robert and Joo, Balint and Lin, Huey-Wen and Orginos, Kostas and
  Richards, David and Thacker, Hank</editor>
  <pubdate>2008</pubdate>
  <volume>LATTICE2008</volume>
  <fpage>175</fpage>
</bibl>

<bibl id="B13">
  <title><p>Deep {Strong} {Coupling} {Regime} of the {Jaynes}-{Cummings}
  {Model}</p></title>
  <aug>
    <au><snm>Casanova</snm><fnm>J.</fnm></au>
    <au><cnm>others</cnm></au>
  </aug>
  <source>Physical Review Letters</source>
  <pubdate>2010</pubdate>
  <volume>105</volume>
  <issue>26</issue>
  <fpage>263603</fpage>
  <url>https://link.aps.org/doi/10.1103/PhysRevLett.105.263603</url>
</bibl>

<bibl id="B14">
  <title><p>Confinement of quarks</p></title>
  <aug>
    <au><snm>Wilson</snm><fnm>KG</fnm></au>
  </aug>
  <source>Physical Review D</source>
  <pubdate>1974</pubdate>
  <volume>10</volume>
  <issue>8</issue>
  <fpage>2445</fpage>
  <lpage>-2459</lpage>
  <url>https://link.aps.org/doi/10.1103/PhysRevD.10.2445</url>
</bibl>

<bibl id="B15">
  <title><p>The anisotropic {Wilson} gauge action</p></title>
  <aug>
    <au><snm>Klassen</snm><fnm>TR</fnm></au>
  </aug>
  <source>Nuclear Physics B</source>
  <pubdate>1998</pubdate>
  <volume>533</volume>
  <issue>1</issue>
  <fpage>557</fpage>
  <lpage>-575</lpage>
  <url>https://www.sciencedirect.com/science/article/pii/S0550321398005100</url>
</bibl>

<bibl id="B16">
  <title><p>Continuum limit and improved action in lattice theories: ({I}).
  {Principles} and $\phi^4$ theory</p></title>
  <aug>
    <au><snm>Symanzik</snm><fnm>K.</fnm></au>
  </aug>
  <source>Nuclear Physics B</source>
  <pubdate>1983</pubdate>
  <volume>226</volume>
  <issue>1</issue>
  <fpage>187</fpage>
  <lpage>-204</lpage>
  <url>https://www.sciencedirect.com/science/article/pii/0550321383904686</url>
</bibl>

<bibl id="B17">
  <title><p>{Instantons from over - improved cooling}</p></title>
  <aug>
    <au><snm>Garcia Perez</snm><fnm>M</fnm></au>
    <au><snm>Gonzalez Arroyo</snm><fnm>A</fnm></au>
    <au><snm>Snippe</snm><fnm>JR</fnm></au>
    <au><snm>Baal</snm><fnm>P</fnm></au>
  </aug>
  <source>Nucl. Phys. B</source>
  <pubdate>1994</pubdate>
  <volume>413</volume>
  <fpage>535</fpage>
  <lpage>-552</lpage>
</bibl>

<bibl id="B18">
  <title><p>{Improved actions for QCD thermodynamics on the
  lattice}</p></title>
  <aug>
    <au><snm>Beinlich</snm><fnm>B.</fnm></au>
    <au><snm>Karsch</snm><fnm>F.</fnm></au>
    <au><snm>Laermann</snm><fnm>E.</fnm></au>
  </aug>
  <source>Nucl. Phys. B</source>
  <pubdate>1996</pubdate>
  <volume>462</volume>
  <fpage>415</fpage>
  <lpage>-436</lpage>
</bibl>

<bibl id="B19">
  <title><p>{Topology of the SU(2) vacuum: A Lattice study using improved
  cooling}</p></title>
  <aug>
    <au><snm>Forcrand</snm><fnm>P</fnm></au>
    <au><snm>Garcia Perez</snm><fnm>M</fnm></au>
    <au><snm>Stamatescu</snm><fnm>IO</fnm></au>
  </aug>
  <source>Nucl. Phys. B</source>
  <pubdate>1997</pubdate>
  <volume>499</volume>
  <fpage>409</fpage>
  <lpage>-449</lpage>
</bibl>

<bibl id="B20">
  <title><p>{Computation of the one loop Symanzik coefficients for the square
  action}</p></title>
  <aug>
    <au><snm>Snippe</snm><fnm>JR</fnm></au>
  </aug>
  <source>Nucl. Phys. B</source>
  <pubdate>1997</pubdate>
  <volume>498</volume>
  <fpage>347</fpage>
  <lpage>-396</lpage>
</bibl>

<bibl id="B21">
  <title><p>{Highly improved lattice field strength tensor}</p></title>
  <aug>
    <au><snm>Bilson Thompson</snm><fnm>SO</fnm></au>
    <au><snm>Leinweber</snm><fnm>DB</fnm></au>
    <au><snm>Williams</snm><fnm>AG</fnm></au>
  </aug>
  <source>Annals Phys.</source>
  <pubdate>2003</pubdate>
  <volume>304</volume>
  <fpage>1</fpage>
  <lpage>-21</lpage>
</bibl>

<bibl id="B22">
  <title><p>{A Novel improved action for SU(3) lattice gauge
  theory}</p></title>
  <aug>
    <au><snm>Langfeld</snm><fnm>K</fnm></au>
  </aug>
  <pubdate>2004</pubdate>
</bibl>

<bibl id="B23">
  <title><p>{Lattice QCD: A Critical status report}</p></title>
  <aug>
    <au><snm>Jansen</snm><fnm>K</fnm></au>
  </aug>
  <source>PoS</source>
  <editor>Aubin, Christopher and Cohen, Saul and Dawson, Chris and Dudek, Jozef
  and Edwards, Robert and Joo, Balint and Lin, Huey-Wen and Orginos, Kostas and
  Richards, David and Thacker, Hank</editor>
  <pubdate>2008</pubdate>
  <volume>LATTICE2008</volume>
  <fpage>010</fpage>
</bibl>

<bibl id="B24">
  <title><p>Review of summation-by-parts schemes for initial--boundary-value
  problems</p></title>
  <aug>
    <au><snm>Sv{\"a}rd</snm><fnm>M</fnm></au>
    <au><snm>Nordstr{\"o}m</snm><fnm>J</fnm></au>
  </aug>
  <source>Journal of Computational Physics</source>
  <publisher>Elsevier</publisher>
  <pubdate>2014</pubdate>
  <volume>268</volume>
  <fpage>17</fpage>
  <lpage>-38</lpage>
</bibl>

<bibl id="B25">
  <title><p>Review of summation-by-parts operators with simultaneous
  approximation terms for the numerical solution of partial differential
  equations</p></title>
  <aug>
    <au><snm>Fern{\'a}ndez</snm><fnm>DCDR</fnm></au>
    <au><snm>Hicken</snm><fnm>JE</fnm></au>
    <au><snm>Zingg</snm><fnm>DW</fnm></au>
  </aug>
  <source>Computers \& Fluids</source>
  <publisher>Elsevier</publisher>
  <pubdate>2014</pubdate>
  <volume>95</volume>
  <fpage>171</fpage>
  <lpage>-196</lpage>
</bibl>

<bibl id="B26">
  <title><p>{Lattice QCD in curved spacetimes}</p></title>
  <aug>
    <au><snm>Yamamoto</snm><fnm>A</fnm></au>
  </aug>
  <source>Phys. Rev. D</source>
  <pubdate>2014</pubdate>
  <volume>90</volume>
  <issue>5</issue>
  <fpage>054510</fpage>
</bibl>

<bibl id="B27">
  <title><p>{Classical Electrodynamics}</p></title>
  <aug>
    <au><snm>Jackson</snm><fnm>JD</fnm></au>
  </aug>
  <publisher>Wiley</publisher>
  <pubdate>1998</pubdate>
</bibl>

<bibl id="B28">
  <title><p>Computational {Electrodynamics}: {The} {Finite}-difference
  {Time}-domain {Method}</p></title>
  <aug>
    <au><snm>Taflove</snm><fnm>A.</fnm></au>
    <au><snm>Hagness</snm><fnm>S.C.</fnm></au>
  </aug>
  <publisher>Artech House</publisher>
  <series><title><p>Artech {House} antennas and propagation
  library</p></title></series>
  <pubdate>2005</pubdate>
  <url>https://books.google.no/books?id=n2ViQgAACAAJ</url>
</bibl>

<bibl id="B29">
  <title><p>On enforcing {Gauss}' law in electromagnetic particle-in-cell
  codes</p></title>
  <aug>
    <au><snm>Bruce Langdon</snm><fnm>A.</fnm></au>
  </aug>
  <source>Computer Physics Communications</source>
  <pubdate>1992</pubdate>
  <volume>70</volume>
  <issue>3</issue>
  <fpage>447</fpage>
  <lpage>-450</lpage>
  <url>https://www.sciencedirect.com/science/article/pii/0010465592901058</url>
</bibl>

<bibl id="B30">
  <title><p>Finite Volume Methods for Hyperbolic Problems</p></title>
  <aug>
    <au><snm>LeVeque</snm><fnm>RJ</fnm></au>
  </aug>
  <publisher>Cambridge University Press</publisher>
  <series><title><p>Cambridge Texts in Applied Mathematics</p></title></series>
  <pubdate>2002</pubdate>
</bibl>

<bibl id="B31">
  <title><p>Effective-field theories for heavy quarkonium</p></title>
  <aug>
    <au><snm>Brambilla</snm><fnm>N</fnm></au>
    <au><snm>Pineda</snm><fnm>A</fnm></au>
    <au><snm>Soto</snm><fnm>J</fnm></au>
    <au><snm>Vairo</snm><fnm>A</fnm></au>
  </aug>
  <source>Reviews of Modern Physics</source>
  <pubdate>2005</pubdate>
  <volume>77</volume>
  <issue>4</issue>
  <fpage>1423</fpage>
  <lpage>-1496</lpage>
  <url>https://link.aps.org/doi/10.1103/RevModPhys.77.1423</url>
</bibl>

<bibl id="B32">
  <title><p>Trace preserving quantum dynamics using a novel
  reparametrization-neutral summation-by-parts difference operator</p></title>
  <aug>
    <au><snm>Ålund</snm><fnm>O</fnm></au>
    <au><cnm>others</cnm></au>
  </aug>
  <source>Journal of Computational Physics</source>
  <pubdate>2021</pubdate>
  <volume>425</volume>
  <fpage>109917</fpage>
  <url>https://www.sciencedirect.com/science/article/pii/S0021999120306914</url>
</bibl>

<bibl id="B33">
  <title><p>Fermion production from real-time lattice gauge theory in the
  classical-statistical regime</p></title>
  <aug>
    <au><snm>Kasper</snm><fnm>V</fnm></au>
    <au><snm>Hebenstreit</snm><fnm>F</fnm></au>
    <au><snm>Berges</snm><fnm>J</fnm></au>
  </aug>
  <source>Physical Review D</source>
  <pubdate>2014</pubdate>
  <volume>90</volume>
  <issue>2</issue>
  <fpage>025016</fpage>
  <url>http://arxiv.org/abs/1403.4849</url>
</bibl>

<bibl id="B34">
  <title><p>Rattle: {A} “velocity” version of the shake algorithm for
  molecular dynamics calculations</p></title>
  <aug>
    <au><snm>Andersen</snm><fnm>HC</fnm></au>
  </aug>
  <source>Journal of Computational Physics</source>
  <pubdate>1983</pubdate>
  <volume>52</volume>
  <issue>1</issue>
  <fpage>24</fpage>
  <lpage>-34</lpage>
  <url>https://www.sciencedirect.com/science/article/pii/0021999183900141</url>
</bibl>

<bibl id="B35">
  <title><p>Finite {Element} {Systems} for vector bundles : elasticity and
  curvature</p></title>
  <aug>
    <au><snm>Christiansen</snm><fnm>SH</fnm></au>
    <au><snm>Hu</snm><fnm>K</fnm></au>
  </aug>
  <source>arXiv:1906.09128 [cs, math]</source>
  <pubdate>2020</pubdate>
  <url>http://arxiv.org/abs/1906.09128</url>
</bibl>

<bibl id="B36">
  <title><p>Discretization of differential geometry for computational gauge
  theory</p></title>
  <aug>
    <au><snm>Schubel</snm><fnm>MD</fnm></au>
  </aug>
  <source>PhD thesis</source>
  <publisher>University of Illinois at Urbana-Champaign</publisher>
  <pubdate>2018</pubdate>
</bibl>

<bibl id="B37">
  <title><p>Second order gauge invariant discretizations to the
  {Schr}{\textbackslash}"odinger and {Pauli} equations</p></title>
  <aug>
    <au><snm>Christiansen</snm><fnm>SH</fnm></au>
    <au><snm>Halvorsen</snm><fnm>TG</fnm></au>
  </aug>
  <source>arXiv:1505.08040 [math]</source>
  <pubdate>2015</pubdate>
  <url>http://arxiv.org/abs/1505.08040</url>
</bibl>

</refgrp>
} 


\end{backmatter}


\end{document}